\documentclass[conference,compsoc]{IEEEtran}
\usepackage{graphicx, subcaption}
\usepackage{algorithm, algpseudocode}
\usepackage{enumitem}
\usepackage{fixltx2e}
\usepackage{booktabs}
\usepackage{amssymb}
\usepackage{multirow}
\usepackage{afterpage}
\begin{document}
\title{Cache-aware data structures for packet forwarding tables on general purpose CPUs}
\author{\IEEEauthorblockN{Maksim Yegorov}
\IEEEauthorblockA{Computer Science Department\\
Rochester Institute of Technology, NY\\ Email: \texttt{mey5634@rit.edu}
}}

\maketitle

\begin{abstract}
Longest prefix matching has long been the bottleneck of the Bloom
filter-based solutions for packet forwarding implemented in software.
We propose a search algorithm to match a destination IP address against a
compact representation of the FIB table in CPU cache on general purpose
hardware with an average performance target of
$\mathcal{O}(\log n)$ for an $n$-bit address.

\end{abstract}

\section{Background}
To set the context, consider the fundamental task of a network router.
In order to match an incoming packet to an outgoing interface link,
the router inspects the packet's header to obtain the destination address.
It will then consult a forwarding table (Forwarding Information Base, 
or FIB) stored on the router.
An address entry in such a table will contain, among other things,
a variable length prefix (as in 129.12.30.0/20).
In effect, the router will compare the destination IP
against the known prefixes using the longest prefix match rule.
The forwarding table may be occasionally updated with new prefixes
received via BGP route advertisement \cite{Kurose:Networking}.

It would appear that a standard hash table or a binary search tree could
satisfy the requirements of the data structure that calls for a
\textit{look-up} and \textit{update} operations. The difficulty with
adapting these fundamental datastructures for Internet routing stems 
mainly from the sheer throughput requirements of today's line speed coupled
with the FIB table size.
This can be illustrated with an example of a hypothetical 50Gbps core router 
and an Ethernet frame of 84 bytes for a minimum sized packet.
The bit per second
wire line speed can be recast in terms of packets per second. Specifically,
for this simplified scenario, the router can be expected to process 75Mpps. 
This may require at least 75 million lookups per second -- or an order of
magnitude more, depending on the implementation of approximate prefix
matching.
For a general purpose CPU clock
speed of (for the sake of example) 4 GHz, this equates to approximately 
50 CPU cycles per packet. If the router cannot keep up with the speed of
arriving packets, it will drop packets.

How much work can be done in 50 CPU cycles? As a very rough approximation,
consider that conventional hashing algorithms require about 10 cycles 
per byte of hash (40 cycles to compute a 32 bit hash). Memory latency
presents a particular challenge. The access times range between 4-50 cycles
for L1 and L3 CPU caches, respectively. The penalty for misses can easily 
double the time requirements. Main memory access will require several hundred 
cycles.

Clearly, this hypothetical scenario is an oversimplication. The forwarding
task is only one among many processing steps that a router performs on each
packet, contemporary CPUs will likely have multiple cores, router line
speeds may be in the single digits or in the hundreds of Gbps, there will 
be a distribution of packet sizes (my estimate errs on the conservative
side), leaf node routers may benefit from caching previously seen IP
addresses etc. Still, the above generalization gives us a ballpark number 
to quickly determine if a particular datastructure is fit for the task.

In view of these numbers, it is not at all surprising that the lookup has 
traditionally been performed in hardware, using dedicated TCAM and SRAM 
circuits. There are multiple considerations that make software 
implementations superior to ASIC hardware based ones. The cost per 
transistor, power requirements, and monopoly effects, in particular, 
drive up the cost. The inability to patch hardware makes security updates
unfeasible. There has been a renewed push recently to develop 
programmable routers that, on the one hand, can accommodate the data 
processing speeds expected of today's networks, and on the other, offer 
the option to implement and continuously update various parts of the 
network stack in software rather than hardware.

\section{Related Work}
Classical algorithms developed up to about 2007 have been surveyed in
\cite{Sanchez:Survey} and \cite{Varghese:Algorithmics}. 
The data structures include trie, tree, and hash table variants.

Of particular relevance is the binary search on prefix lengths
proposed in \cite{Waldvogel:Scalable}. Waldvogel et al. propose a very 
elaborate hash table of binary search trees with logarithmic time 
complexity. Most
of the refinements involve comparatively large databases that require at
least an order of magnitude more memory than what can fit into third level
cache, and are therefore only practical for hardware implementations. We
believe that the core ideas of leveraging the binary search on prefixes and 
using memoization to avoid backtracking can be adapted for the more 
compact data structures.

Dharmapurikar et al.\cite{Dharmapurikar:Bloom} describe a longest prefix
matching algorithm utilizing a probabilistic set membership check with
Bloom filters. A Bloom filter is associated with each prefix length. The
destination adress is masked and matched against each of the Bloom filters,
yielding a list of one or more prefix matches. The list is then checked
against an off-chip conventional hash table, starting with the longest 
prefix match. Because of the arbitrarily small false positive rate, 
a single lookup in high-latency main memory is sufficient in practice.

\section{Solution}

\subsection{Goals}

We have identified two opportunities for improvement in the context of
the Bloom filter-based solutions to the longest prefix matching problem.
The Bloom filter (BF) data structure was originally used by 
Dharmapurikar et al. \cite{Dharmapurikar:Bloom}
for parallel look up implemented in hardware. By contrast, the software
implementations on conventional hardware pay a hefty penalty -- in computation
cost and code complexity -- to parallelize the look up. Consequently, linear
search has been the default solution to the longest prefix matching problem
(see Algorithm~\ref{alg:linearsearch}). The time
complexity of Algorithm~\ref{alg:linearsearch} is $\mathcal{O}(n)$, where $n$
is the number of distinct prefix lengths in the BF. We
propose to improve on linear search for Bloom filter in this paper.

\begin{algorithm}
\caption{Linear search for longest matching prefix}\label{alg:linearsearch}
\begin{algorithmic}[1]
\Procedure{LinearSearch}{$bf, ip, fib, maxlen$}
\State $plen \gets maxlen$\Comment{max prefix length}
\While{$plen \geq \textsc{minlen}$}\Comment{min prefix length}

  \State $tmp \gets$ extract $plen$ most significant bits in IP
  \State $key \gets \texttt{encode(tmp, plen)}$
  \State $res \gets \texttt{bf.contains(key,}$
                \State $\hspace*{36mm} \texttt{[hash\textsubscript{1}..hash\textsubscript{bf.k}])}$
  \If{$res \not= 0 \;\&\; key \in fib$}
      \State \textbf{return} $plen$
  \Else
    $\;plen \gets plen - 1$
  \EndIf
\EndWhile
\State \textbf{return} \textsc{pref\textsubscript{default}}\Comment{default route}
\EndProcedure
\end{algorithmic}
\end{algorithm}

Second, any scheme that utilizes a probabilistic data structure, such as
the Bloom filter, to identify candidate(s) for the \emph{longest matching
prefix} (LMP) would generally need to look up the candidate(s)
in a forwarding table that serves as the definitive membership
test and the store of the next hop information. Current solutions typically
store this information in an off-chip hash table. This operation is therefore
a bottleneck of the probabilistic filter-based schemes. We conjecture that
the method we propose is broadly applicable to any key-value store
application that

\begin{enumerate}[label=(\alph*)]
\item is defined as a many-to-few kind of mapping over totally ordered keys, and
\item tolerates (i.e. self-corrects for) a certain probability of error.
\end{enumerate}

In the case of the FIB table, we propose to store the outgoing link information
in a compact array. We would then insert encoded \texttt{(index, prefix)} pairs
into a \emph{guided BF} data structure (\texttt{BF\textsubscript{fib}}, 
separate from \texttt{BF\textsubscript{lmp}} used to
encode \texttt{(length, prefix)} pairs). This scheme is feasible for
forwarding table applications on today's off-the-shelf hardware with typical
requirements of on the order of a million prefix keys and outgoing
interfaces numbering in the low hundreds.

From our preliminary analysis, both Bloom filters (\texttt{BF\textsubscript{fib}}
and \texttt{BF\textsubscript{lmp}}) can fit in L3 cache assuming
current backbone router FIB table sizes that we have
surveyed.\cite{Yegorov:github} A low-level implementation and a cost-benefit
analysis of such a FIB representation are in progress.

\subsection{Implementation}

\begin{figure}[h]
\centering
\includegraphics[height=2.6in]{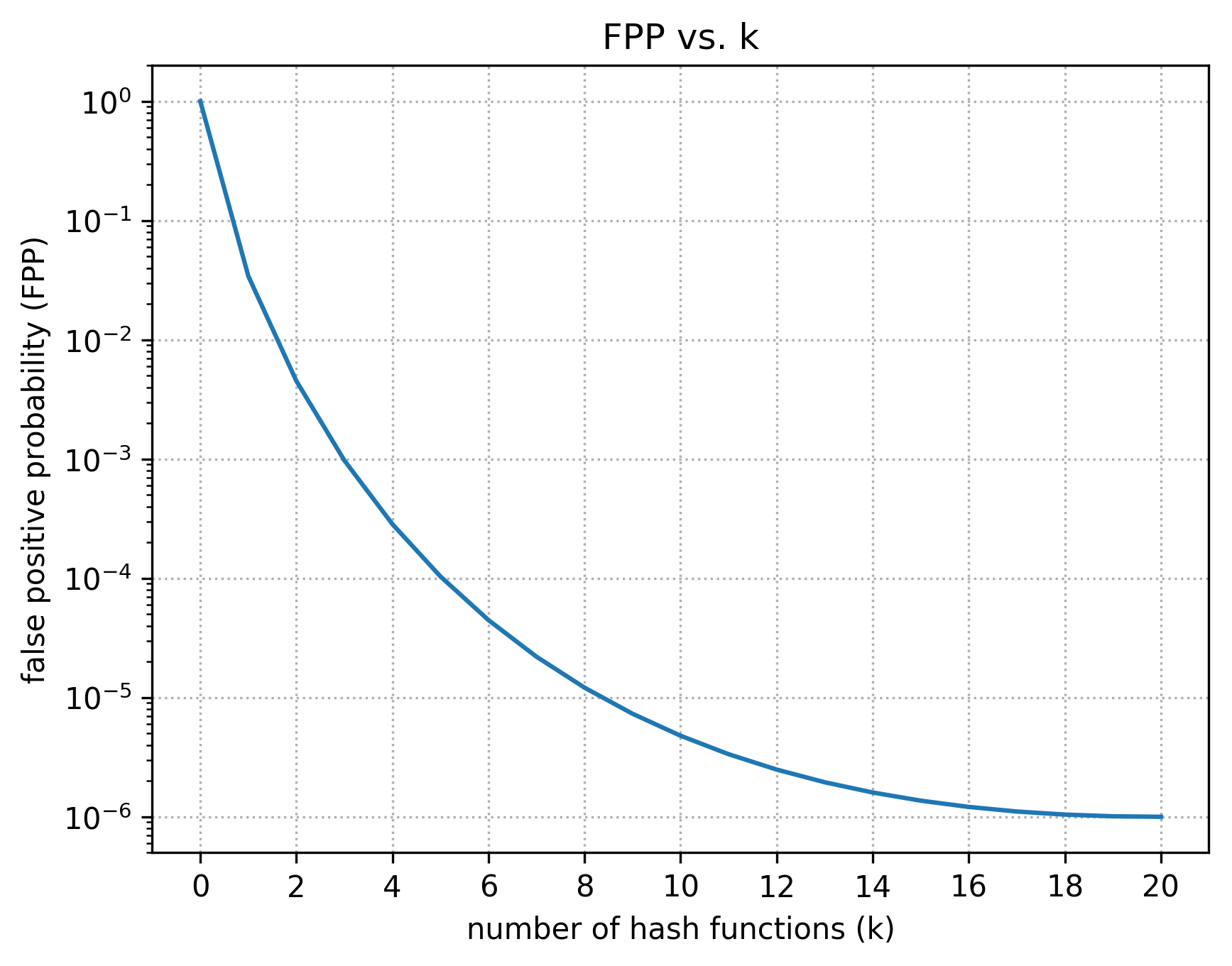}
  \caption{False positive probability vs. number of hash functions in an \emph{optimal} BF}
\label{fig:fpp}
\end{figure}

The key observation that we draw upon is that any one of the routine
tests -- whether a particular bit in the Bloom filter bit vector
is set -- contains valuable information, in that the correlation between
a set bit and a prefix being a member of the set is much higher than chance
(see Fig.~\ref{fig:fpp}). The cost of calculations performed as part of
validating the membership of a given key in a BF gives us an incentive to
assign meaning to specific hashing calculations. In other words, we will 
define a simple protocol that exploits the overhead associated with the
BF hashing calculations to direct the search for the longest matching prefix.

Algorithm~\ref{alg:build} contains the pseudocode to \emph{insert} a prefix
(\emph{pref}) into a BF (\emph{bf}) that lends itself to 
\emph{guided search}. The goal is to pre-compute the search 
path for an IP that could possibly match a prefix in the BF
while the data structure is built up.
Algorithm~\ref{alg:guidedsearch} suggests
a procedure to \emph{look up} an IP in the BF constructed by
Algorithm~\ref{alg:build}. The
algorithms assign specific meaning to the \emph{first} hashing function
to direct our search left or right in a binary search tree. In addition, we
reserve $n$ hashing functions ($n\in\{5,7\}$ suffice for the IPv4 and IPv6
tables used in our experiments) to encode
the best matching prefix as a bit sequence. The $n$ bits, when decoded,
will index the prefix length in a compact array of distinct prefix lengths
found in the router's FIB table (e.g., for the IPv4 table used in our 
experiments, $\texttt{ix 0}$ $\rightarrow$ 
$\texttt{pref len 0}$, $\texttt{ix 1}$ $\rightarrow$ $\texttt{pref len 8}$
etc.).

Both the \emph{build} and the \emph{look up} procedures assume a
binary search tree (\emph{bst}) to guide the search. An approximately 
optimal tree could
be constructed for a given router if the historic traffic data and its 
correlation with the fraction of the address space covered by each prefix length
were known. In the absence of such information, we can conservatively
assume random traffic and a balanced binary search tree (as in the
classic binary search algorithm).

The \emph{build} invokes the \emph{look up} to identify the
best matching (shorter) prefix in the BF constructed to date for the 
(longer) prefix
about to be inserted. Accordingly, we sort the prefixes before
inserting them into the BF in the ascending order.

\begin{algorithm}
  \caption{Build a BF to enable guided search for LMP}\label{alg:build}
  \begin{algorithmic}[1]

    \Procedure{Insert}{$bf, pref, fib, bst$}

    \State $bmp \gets \texttt{Lookup(}$\Comment{best match prefix}
        \State $\hspace*{20mm} \texttt{bf,}$
        \State $\hspace*{20mm} \texttt{pref,}$
        \State $\hspace*{20mm} \texttt{fib,}$
        \State $\hspace*{20mm} \texttt{bst)}$
    \State $curr \gets bst$\Comment{start at root}
    \State $count_{hit} \gets 0$\Comment{times branched right}

    \While{$curr \not= null$}\Comment{not leaf}

      \If{$pref.len < curr.plen$}
        \State $curr \gets curr.left$
      \ElsIf{$pref.len = curr.plen$}
        \State $key \gets \texttt{encode(pref, pref.len)}$
        \State $\texttt{bf.ins(key,}$
                \State $\hspace*{15mm} \texttt{[hash\textsubscript{1}..hash\textsubscript{bf.k}])}$
        \State \textbf{break}
      \Else
        \State $tmp \gets$ $curr.plen$ most signif bits in $pref$
        \State $key \gets \texttt{encode(tmp, curr.plen)}$
        \State $\texttt{bf.ins(key, [hash\textsubscript{1}])}$\Comment{signal right}
        \State $count_{hit} \gets count_{hit} + 1$
        \State $hashes \gets \texttt{filter(}$\Comment{hash funcs}
              \State $\hspace*{15mm} \texttt{bmp,}$\Comment{encode bmp}
              \State $\hspace*{15mm} \texttt{hash\textsubscript{count\textsubscript{hit}},}$\Comment{start hash func}
              \State $\hspace*{15mm} \texttt{n)}$\Comment{num bits}
        \State $\texttt{bf.ins(key, hashes)}$
        \State $curr \gets curr.right$
      \EndIf
    \EndWhile
    \EndProcedure

  \end{algorithmic}
\end{algorithm}

Algorithm~\ref{alg:guidedsearch} defaults to linear search when a bit that
would be unset
under the perfect hashing assumption \emph{is} set in the actual BF.
The \emph{guided} search can reach a dead end when

\begin{enumerate}
  \item the first hashing function directs it \emph{right}, where
    (in hindsight) it should have pointed \emph{left};
  \item the decoded best matching prefix length is incorrect -- either
    logically impossible or failing the BF look up on one of the
    remaining hash functions;
  \item the case of false positive: BF contains a prefix not found in
    FIB.
\end{enumerate}

In any one of these cases, a stalemate is avoided by defaulting to the 
linear look up scheme (Algorithm~\ref{alg:linearsearch}), starting
just below the longest match to date ($last_{hit}$).

Given the number of prefixes to be stored in the BF, we can tune the BF 
parameters (bit array size $m$, number of hash functions $k$) to provide
an optimal balance between the size of the data structure in memory (i.e.,
design the BF to fit in CPU cache), on the one hand, and the rate at which
the guided search would default to linear search and the FIB look up 
rate, on the other. The cost benefit analysis is a
function of the available L3 cache size , the penalty for off-chip memory 
hits and misses, the computational cost per byte of hash, and the like --
and can be established through grid search and tuned for the target 
hardware (and traffic, if the details are available).

\begin{algorithm}
  \caption{Guided search for LMP}\label{alg:guidedsearch}
  \begin{algorithmic}[1]

    \Procedure{Lookup}{$bf, ip, fib, bst$}
      \State $last_{hit} \gets -1$\Comment{last plen that yielded hit}
      \State $count_{hit} \gets 0$\Comment{times branched right}
      \State $curr \gets bst$\Comment{start at root}

      \While{$curr \not= null$}\Comment{not leaf}
        \State $tmp \gets$ $curr.plen$ most significant bits of IP
        \State $key \gets \texttt{encode(tmp, curr.plen)}$
        \State $res \gets \texttt{bf.contains(tmp, [hash\textsubscript{1}])}$
        \If{$res = 1$}
          \State $count_{hit} \gets count_{hit} + 1$
          \State $last_{hit} \gets curr.plen$
          \State $curr \gets curr.right$
        \Else
          \State $curr \gets curr.left$
        \EndIf
      \EndWhile\Comment{reached leaf}

      \If{$last_{hit}=-1$}
        \State \textbf{return} \textsc{pref\textsubscript{default}}\Comment{default route}
      \EndIf

      \State $tmp \gets last_{hit}$ most significant bits of IP
      \State $key \gets \texttt{encode(tmp, last\textsubscript{hit})}$
      \State $bmp \gets \texttt{bf.contains(}$\Comment{decode best match}
              \State $\hspace*{25mm} \texttt{key,}$
              \State $\hspace*{25mm} \texttt{[hash\textsubscript{count\textsubscript{hit}}..hash\textsubscript{count\textsubscript{hit}+n-1}],}$
              \State $\hspace*{25mm} \texttt{decode=true)}$

      \If{$bmp=2^n-1 \;|\; bmp < last_{hit}$}
        \State $key \gets$ encode best match prefix, as usual
        \State $res \gets \texttt{bf.contains(}$
              \State $\hspace*{25mm} \texttt{key,}$
              \State $\hspace*{25mm} \texttt{[hash\textsubscript{count\textsubscript{hit}+n}..hash\textsubscript{bf.k}])}$

        \If{$res \neq 0 \;\&\; key \in fib$}
            \State \textbf{return} $bmp$
        \EndIf
      \EndIf

      \State \textbf{return} \texttt{LinearSearch(}\Comment{defaulting}
      \State $\hspace*{25mm} \texttt{bf, ip,}$
      \State $\hspace*{25mm} \texttt{fib, last\textsubscript{hit}-1)}$
    \EndProcedure

  \end{algorithmic}
\end{algorithm}

Because of the possibility of defaulting to linear search, the time
complexity of Algorithm~\ref{alg:guidedsearch} is $\Omega(\log n)$, 
where $n$ is the number of distinct prefix lengths in the BF.
The BF parameters can be chosen to control the default rate for
$\mathcal{O}(\log n)$ average case performance, in the same
way as the false positive rate can be tuned for the standard BF.
In practice, the degree to which the default rate can be minimized is
limited by the practical considerations of the available CPU cache size.

In summary, for each packet, the \emph{guided} search scales the full
height of the binary search tree until it reaches a leaf, then decodes
the best matching prefix (\emph{bmp}) from the most recent
\texttt{hash\textsubscript{1}} match, and finally verifies
the match using remaining hash functions on the \emph{bmp} itself.
Occasionally, it will default to linear search over the lower prefix 
lengths.

\section{Experiments}

\subsection{Design of Experiments}

Table~\ref{tab:experiment-matrix} summarizes the experiments that we have
run. The goal has been to compare the performance of the linear and guided 
search schemes in terms that

\begin{enumerate}[label=(\alph*)]
\item are common to both algorithms, and
\item account for the bulk of CPU and memory access time, irrespective of 
  implementation.
\end{enumerate}

In particular, any filter-based implementation will involve repeated testing
if a given bit is set in the filter's bit vector, invoking a
non-cryptographic hash function, and looking up a candidate prefix match in
FIB (whether implemented using BF or hash table). In the case
of the guided BF search, we may also be interested to know, how frequently
the BF (with given parameter settings) defaults to linear search. Obviously,
the default rate will also be reflected in the other metrics.
With this in mind, we instrumented the BF implementation to collect the
statistics on bit lookup, hashing, and FIB table lookup function
invocations. We report profiling results per packet.

The core router BGP data is obtained from the University of Oregon Route
Views Project.\cite{Oregon:RouteViews}
The performance of either search scheme will depend on the traffic passing
through the router in question. The traffic may be more or less correlated
with the prefixes in the table. We benchmark both search schemes on three
synthetically produced traffic data sets:

\begin{enumerate}
\item Random traffic: IP addresses are chosen randomly from the
    address space. Because of the vast size of the IPv6 address space,
    the absolute majority of randomly selected addresses match
    to the default (length \texttt{/0}) route.
\item Traffic is generated randomly from the address space spanned by the
    prefixes in the table in proportion to the fraction of the address space
    covered by the prefixes of a given length. For example, given an IPv4
    BGP table where 16-bit long prefixes cover 1/4 of the total address
    space, we use reservoir sampling to generate an address from the
    the set of subnets defined by the length \texttt{/16} prefixes
    in the table with 25\% probability.
\item Traffic is correlated with the frequency distribution of the prefix
    lengths in the BGP table. For example, given an IPv4 table where
    24-bit long prefixes account for 60\% of all prefixes, we use reservoir
    sampling to generate an address spanned by the length \texttt{/24} 
    subnets with 60\% probability.
\end{enumerate}

If the correlation of the traffic passing through a given device with
the prefix distribution in the table were known, we would be able to 
customize the binary search tree and possibly reduce the search time
ammortized over aggregate traffic.
This would involve solving for the optimal binary search tree
by assigning differential weights to each prefix length, so as to reduce the
height of the frequently traveled branches, at the expense of the less
well traveled paths. The optimal tree can then be computed using, for
example, Knuth's dynamic programming algorithm. The
weights are a linear combination of two factors, namely
the fraction of traffic matched to each prefix length and the
height balance ratio among tree branches optimized to avoid gross
imbalance in a scheme where each traversal scales the full height of some
branch from root to leaf.

\begin{figure}[h]
\centering
\includegraphics[height=1.45in]{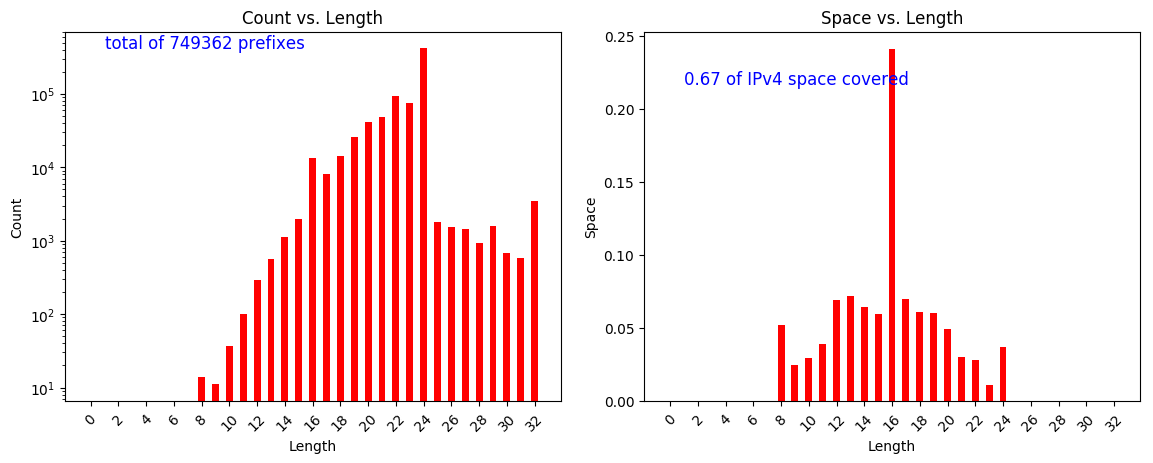}
  \caption{Prefix distribution in the U. of Oregon Route Views IPv4 table}
\label{fig:prefixes-v4}
\end{figure}

\begin{figure}[h]
\centering
\includegraphics[height=1.45in]{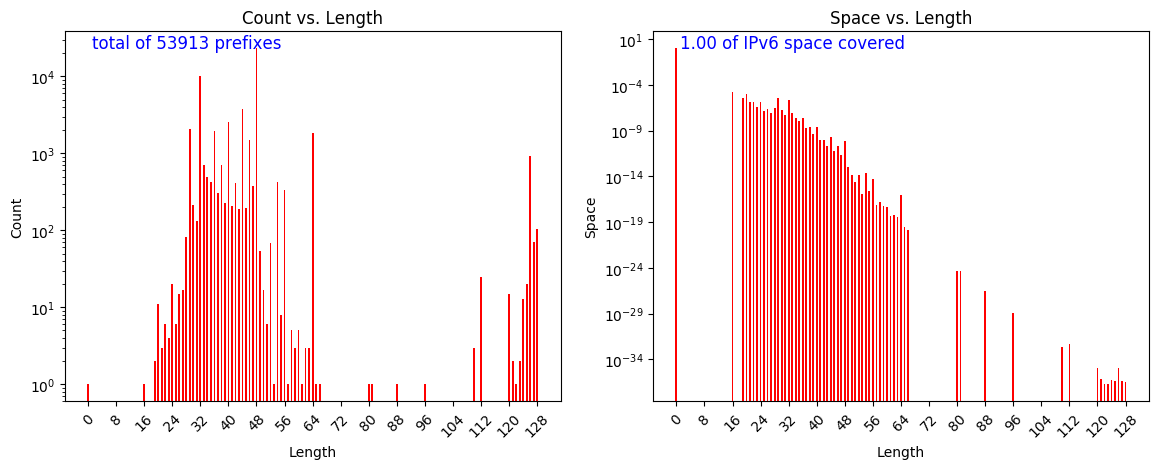}
  \caption{Prefix distribution in the U. of Oregon Route Views IPv6 table}
\label{fig:prefixes-v6}
\end{figure}

In the absence of data on traffic patterns, we merely observe the effect 
of each synthetically generated pattern on the relative performance of the 
guided vs. linear search schemes. In all
experiments, we use the balanced binary tree (equivalently, binary search)
with the implication that prefix lengths in any one branch of the
balanced tree are equally likely to match any one IP.

\begin{figure}[h]
\centering
\includegraphics[height=1.9in]{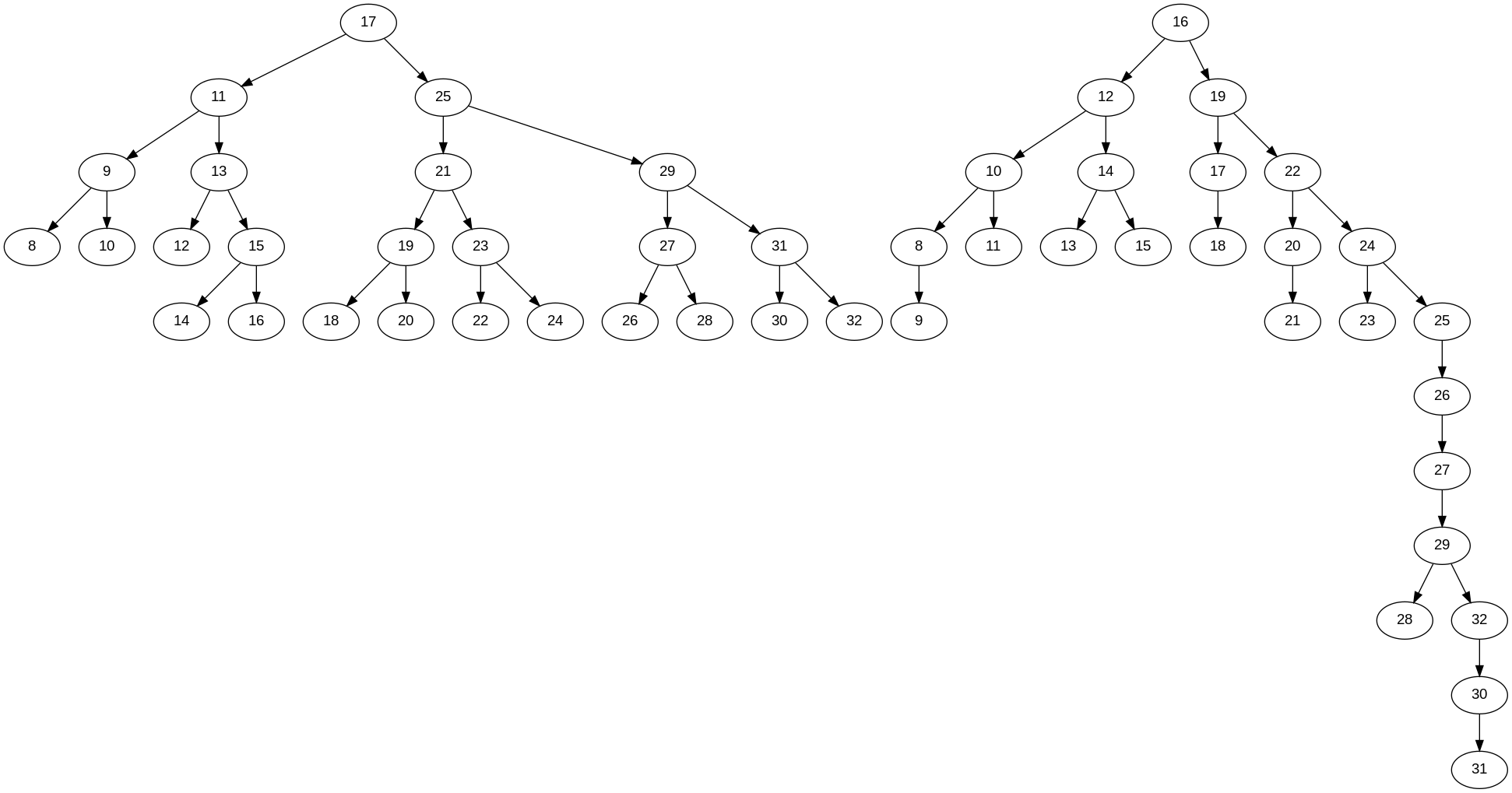}
  \caption{Balanced BST (used in experiments) vs. BST optimized exclusively for address share (not used), both based on the U. of Oregon Route Views IPv4 table}
\label{fig:trees-v4}
\end{figure}

We also contrast the relative performance of the linear and guided search
on IPv4 vs. IPv6 traffic. We use two traffic pattern extremes:

\begin{enumerate}
\item random traffic, where most if not all IPv6 queries go to default route;
\item traffic correlated with the frequency distribution of each prefix 
  length (in the limit, using prefixes themselves as traffic).
\end{enumerate}

Finally, we observe the effect of the BF hyperparameters:
the bit vector size ($m$) -- or equivalently, the percentage of bits set -- 
and the count of hash functions ($k$) on the relative performance of the 
linear vs. guided search schemes.

\begin{table}[]
\centering
\caption{Experiment Matrix} 
\label{tab:experiment-matrix}

\begin{tabular}{@{}lclcl@{}}
\toprule
traffic                 & \multicolumn{4}{c}{protocol}                        \\ \midrule
                        & \multicolumn{2}{c}{IPv4} & \multicolumn{2}{c}{IPv6} \\
random                  & \multicolumn{2}{c}{\checkmark}    & \multicolumn{2}{c}{\checkmark}    \\
by prefix address space & \multicolumn{2}{c}{\checkmark}    & \multicolumn{2}{c}{}     \\
by prefix frequency     & \multicolumn{2}{c}{\checkmark}    & \multicolumn{2}{c}{\checkmark}    \\ \bottomrule

\end{tabular}
\end{table}

\subsection{Discussion}

We summarize the results of the experiments with different IPv4 traffic
types in Table~\ref{tab:traffic}. For IPv4, the \emph{guided search}
requires about one half the number of accesses on a per packet basis compared 
to \emph{linear search}.

\begin{table}[]
\centering
\caption{IPv4 linear vs. guided search performance by traffic type; guided BF: $n=749362$ elements, $k=10$ hash funcs, $size=2.57$MB, $length=21548036$ bits, $33.3$\% full;
linear BF: $fpp=10^{-4}$, $n=749362$ elements, $k=14$ hash funcs, $size=1.71$MB, $length=14365358$ bits, $51.8$\% full}
\label{tab:traffic}

\begin{tabular}{llccl}
\hline
\textbf{traffic}                         & \textbf{metric (per packet)} & \textbf{linear}          & \multicolumn{2}{l}{\textbf{guided}} \\ \hline
\multirow{2}{*}{random}                  & bit lookup      & 49.3                     & \multicolumn{2}{c}{22.1}            \\ 
                                         & hashing         & 20.0                     & \multicolumn{2}{c}{11.2}            \\ \hline
\multirow{2}{*}{by prefix address space} & bit lookup      & 48.0                     & \multicolumn{2}{c}{22.3}            \\ 
                                         & hashing         & 17.4                     & \multicolumn{2}{c}{10.3}            \\ \hline
\multirow{2}{*}{by prefix frequency}     & bit lookup      & 33.3                     & \multicolumn{2}{c}{17.7}            \\ 
                                         & hashing         & 10.3                     & \multicolumn{2}{c}{8.6}             \\ \hline
\end{tabular}
\end{table}

\begin{figure}[h]
\centering
\includegraphics[height=2.2in]{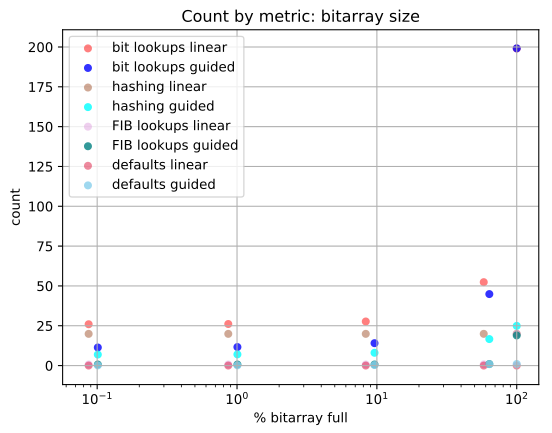}
  \caption{IPv4 search performance for guided vs. linear search; $n=749362$ elements, $k=10$ hash funcs}
\label{fig:utilization}
\end{figure}

Search performance vs. utilization ratio (\% full) is documented in
Table~\ref{tab:utilization}. From Figure~\ref{fig:utilization} it can be
seen, that the linear vs. guided search relative performance levels off 
when the bitarray is approximately 10\% full. 
For a table of approx. 750,000 entries,
10\% full guided BF requires 10.3MB and can therefore fit in
L3 cache on today's general purpose CPUs. The performance for a
compact 2.57MB guided BF is still an improvement over linear search, even 
though the search defaults to linear 68\% of the time. 
The high default rate is
reflected in the increased number of bit lookups and hash computations.

\begin{table}[]
\centering
\caption{IPv4 guided search performance by utilization ratio (random traffic); guided BF: $n=749362$ elements, $k=10$ hash funcs}
\label{tab:utilization}

\begin{tabular}{llcl}
\hline
\textbf{\% bit vector full (size)}              & \textbf{metric} & \multicolumn{2}{l}{\textbf{count per packet}} \\ \hline
\multirow{4}{*}{33.3\% full (2.6MB)} & bit lookup      & \multicolumn{2}{c}{22.1}                           \\
                                     & hashing         & \multicolumn{2}{c}{11.2}                           \\
                                     & FIB lookup      & \multicolumn{2}{c}{0.7}                            \\
                                     & default rate    & \multicolumn{2}{c}{68\%}                           \\ \hline
\multirow{4}{*}{9.6\% full (10.3MB)} & bit lookup      & \multicolumn{2}{c}{14.0}                           \\
                                     & hashing         & \multicolumn{2}{c}{8.0}                            \\
                                     & FIB lookup      & \multicolumn{2}{c}{0.7}                            \\
                                     & default rate    & \multicolumn{2}{c}{30\%}                           \\ \hline
\multirow{4}{*}{4.0\% full (25.7MB)} & bit lookup      & \multicolumn{2}{c}{12.5}                           \\
                                     & hashing         & \multicolumn{2}{c}{7.4}                            \\
                                     & FIB lookup      & \multicolumn{2}{c}{0.7}                            \\
                                     & default rate    & \multicolumn{2}{c}{22\%}                           \\ \hline
\end{tabular}
\end{table}

For IPv4, the number of bit lookups asymptotically approaches approximately 
eight per packet.
With prefixes in the table covering
approximately 67\% of the address space, 33\% of lookups will
require four lookups to arrive at the default route, while the remaining
67\% of lookups will
traverse the full height of the balanced tree (five lookups), in addition to
decoding the 5-bit \emph{best matching prefix} sequence for a total of
ten lookups. This calculation also suggests that $k=10$ hash functions
is the minimum for the protocol to function.

Using the same assumption that the bit vector is sparse enough to never
default to linear search, there would be just over five hash lookups per
packet in the 
limit. Four hash computations are required to yield the default route for
33\% of lookups and six hash computations are required
for the remaining 67\% of lookups, which includes five hashes to traverse 
the height of the tree and a single hash computation to decode 
the 5-bit sequence.

\begin{figure}[p]
\centering
\includegraphics[height=1.5in]{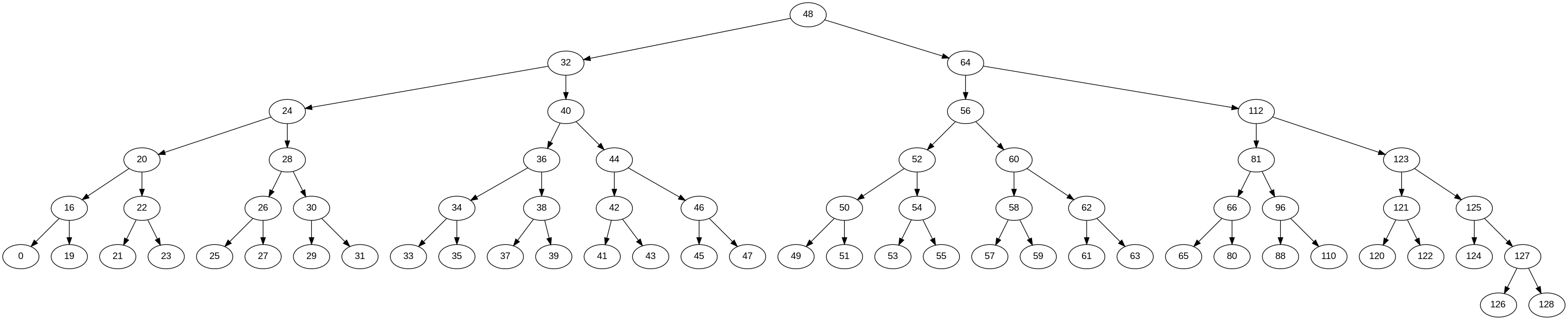}
  \caption{Binary search tree used for the IPv6 prefix search}
\label{fig:IPv6-tree}
\end{figure}

Last but not least, the proposed search algorithm is particularly effective
for the IPv6 address space. This is to be expected with the 128-bit IP
addresses that require only six bit vector lookups to match the default
route with the logarithmic time complexity (Figure~\ref{fig:IPv6-tree}).

\begin{figure}[p]
\centering
\includegraphics[height=2.5in]{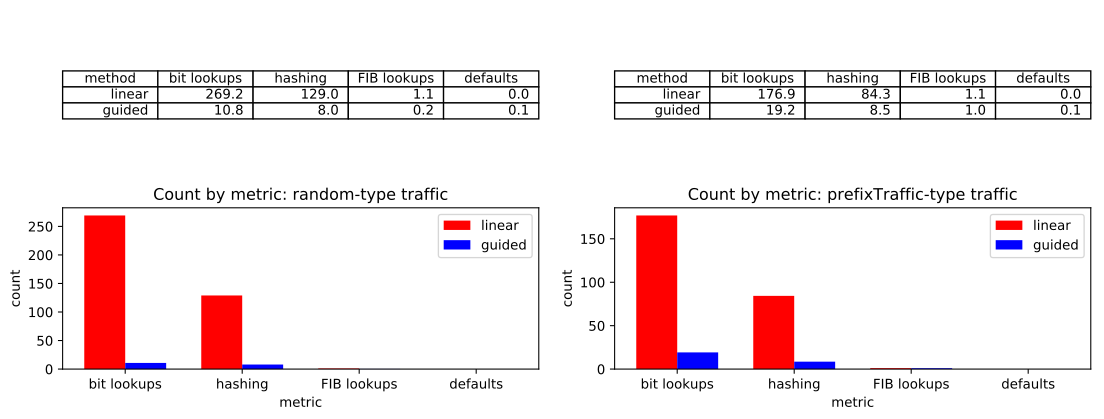}
\caption{IPv6 search performance: random traffic that always defaults to 0-prefix length vs. traffic correlated with the share of each prefix length (no defaults);
  guided BF: $n=53913$ elements, $k=14$ hash funcs, $size=2.05$MB, $length=17238428$ bits, $4.4$\% full;
  linear BF: $fpp=10^{-3}$, $n=53913$ elements, $k=10$, $size=0.09$MB, $length=775139$ bits, $50.1$\% full}
\label{fig:IPv6-by-traffic}
\end{figure}

Figure~\ref{fig:IPv6-by-traffic} illustrates the relative performance of
the two search algorithms for two IPv6 traffic patterns. 
At one extreme, there is the default-only
traffic that is optimally suited for the guided search scheme. At the other
extreme, there is the case of no-default traffic in direct proportion to
the share of each prefix length (here we reuse prefixes as the traffic).
For either traffic pattern, the guided scheme outperforms the linear 
search by an order of magnitude.

\bibliographystyle{IEEEtran}
\bibliography{bibliography}

\clearpage

\clearpage

\end{document}